\documentclass[twocolumn,twoside,slac_two,floatfix]{revtex4}
\usepackage{graphicx}
\usepackage[hypertex]{hyperref}

\usepackage{fancyhdr}
\def\OMIT#1{}
\def\lsim{\lesssim}
\def\gsim{\gtrsim}

\newcommand{\beq}{\begin{equation}}
\newcommand{\eeq}{\end{equation}}
\newcommand{\beqa}{\begin{eqnarray}}
\newcommand{\eeqa}{\end{eqnarray}}
\def\nn{\nonumber}

\def\d{{\rm d}}
\def\ds{\displaystyle}

\newcommand{\Bbar}{\,\overline{\!B}{}}
\newcommand{\Dbar}{\,\overline{\!D}{}}
\newcommand{\Kbar}{\,\overline{\!K}{}}
\def\B0bar{\Bbar{}^0}
\def\D0bar{\Dbar{}^0}
\def\K0bar{\Kbar{}^0}
\def\P0bar{\,\overline{\!P}{}^0}
\def\TeV{{\rm TeV}}
\def\GeV{{\rm GeV}}

\begin{document}

\title{\boldmath Mixing and $CP$ violation in the $D^0$ and $B_s^0$ systems}

\author{Zoltan Ligeti}

\affiliation{Ernest Orlando Lawrence Berkeley National Laboratory,
University of California, Berkeley, CA 94720}

\begin{abstract}

Recent developments for mixing and $CP$ violation in the $D^0$\,--\,$\D0bar$ and
$B_s^0$\,--\,$\B0bar_s$ systems are reviewed, including (i)~the recently
emerging evidence for $D^0$\,--\,$\D0bar$ mixing and the interpretations of the
measurements; (ii) the theoretical status of the calculations of
$\Delta\Gamma_D$ and $\Delta m_D$; (iii) some implications of the measurement of
$B_s^0$\,--\,$\B0bar_s$ mixing for new physics.

\end{abstract}

\maketitle

\thispagestyle{fancy}

\section{Introduction}

Neutral meson mixing provides excellent tests of the Standard Model (SM) and
probes of new physics (NP): $CP$ violation involving $K^0$\,--\,$\K0bar$ mixing
($\epsilon_K$) predicted the third generation; $\Delta m_K$ predicted the charm
mass; $\Delta m_B$ predicted the top mass to be heavy.  While 31 years passed
between the discovery of the $K_L$ (1956) and the discovery of
$B^0$\,--\,$\B0bar$ mixing (1987), after 19 years, in 2006, the
$B_s^0$\,--\,$\B0bar_s$ mixing frequency was measured~\cite{Abulencia:2006ze}
and now the observation of $D^0$\,--\,$\D0bar$
mixing~\cite{Aubert:2007wf,Staric:2007dt} is on the verge of being well
established.  This talk focuses on the implications of these last two sets of
measurements.

Almost all extensions of the SM aimed at solving the hierarchy problem also
contain new sources of $CP$ violation and flavor conversion.  If there is NP at
the TeV scale, flavor physics already imposes strong constraints on it.  Generic
TeV-scale NP models violate the experimental bounds from $K$ and $B$ mixing and
flavor-changing neutral current (FCNC) decay measurements by several orders of
magnitude.  Thus, new flavor physics has to either (i) originate at a much
higher scale than 1\,TeV and be decoupled; or (ii) originate from electroweak
symmetry breaking (EWSB) related NP with non-trivial
structure~\cite{michele,ben}.

Many models with TeV-scale new particles could have given rise to significant
deviations from the SM predictions for $B_s^0$ mixing.  For example, due to its
large mass, the top quark may couple strongly to the NP sector, and in some
scenarios it affects $B_s^0$ mixing, but not $B^0$ or $K$
mixing~\cite{michele,Agashe:2005hk}.  Large $D^0$ mixing is predicted by
quark-squark alignment models~\cite{Nir:1993mx}, since in order not to violate
the $\Delta m_K$ bound, Cabibbo mixing must mostly come from the up sector,
predicting $\Delta m/\Gamma \sim {\cal O}(\lambda^2)$ if $m_{\tilde g, \tilde q}
\lsim 1\,\TeV$.

\subsection{Formalism}

The time evolution of the two flavor eigenstates is
\beq
i\, {{\rm d}\over{\rm d}t} 
  \pmatrix{|P^0(t)\rangle\cr |\P0bar(t)\rangle} 
= \Big(M - {i\over2}\,\Gamma\Big)\! 
  \pmatrix{|P^0(t)\rangle\cr |\P0bar(t)\rangle} \,,
\eeq
where $M$ and $\Gamma$ are $2\times2$ Hermitian matrices, and $CPT$ invariance
implies $M_{11} = M_{22}$ and $\Gamma_{11} = \Gamma_{22}$.
The physical states are eigenvectors of the Hamiltonian,
\beq
|P_{L,H}\rangle = p\, |P^0\rangle \pm q\, |\P0bar\rangle\,.
\eeq
The time evolutions of these heavier ($H$) and lighter ($L$) mass eigenstates
involve  mixing and decay
\beq
|P_{L,H}(t)\rangle = e^{-(im_{L,H}
  + \Gamma_{L,H}/2)t}\,|P_{L,H}\rangle\,.
\eeq
We define the average mass and width by
\beq
m = {m_H+m_L \over 2}\,, \qquad 
  \Gamma = {\Gamma_H+\Gamma_L \over 2}\,,
\eeq
and the mass and width differences
\beq
\Delta m = m_H-m_L\,, \qquad 
  \Delta\Gamma = \Gamma_H - \Gamma_L\,.
\eeq
Note that $\Delta m$ is positive by definition, and the sign of $\Delta\Gamma$
is opposite from the one used by the Tevatron experiments for $B_s^0$.
We denote the decay amplitudes to a final state $f$ by
\beq
A_f = \langle f | {\cal H} |P^0\rangle\,, \qquad
  \bar A_f = \langle f | {\cal H} |\P0bar\rangle\,.
\eeq

Of the there phase-convention independent physical observables,
\beq
\bigg|{\bar A_{\bar f}\over A_f}\bigg| , \qquad
\bigg|{q\over p}\bigg| , \qquad
\lambda_f = {q\over p}\, {\bar A_f\over A_f}\,,
\eeq
deviations of the first two from unity characterize $CP$ violation in decay and
in mixing, respectively, while ${\rm Im}\,\lambda_f \neq 0$ is $CP$ violation in
the interference between decay with and without mixing.  Other phase-convention
independent quantities are
\beq\label{cpobs}
\phi_{12} = {\rm arg}\bigg(\!\!-\frac{M_{12}}{\Gamma_{12}}\bigg)\,, \qquad
{\rm Im}\, {\Gamma_{12}\over M_{12}} = {1-|q/p|^4\over 1+|q/p|^4}\,,
\eeq
where $\phi_{12}$ can easily be modified by NP contributions to $M_{12}$ (this
definition is such that in the SM $\phi_{12}$ is near 0 in the $B_{d,s}$ and $K$
systems).  Unlike $\phi_{12}$, $\arg\, (q/p)$ is phase-convention dependent. 
The second quantity in Eq.~(\ref{cpobs}) --- also known as the dilepton
asymmetry, $A_{\rm SL}$, in $B$ decays, or $-A_m$ in $D$ decays if $|q/p|\approx
1$ --- is subject to hadronic uncertainties.  It is essentially incalculable in
the $D$ and $K$ systems, and its calculation for $B_{d,s}$ using the operator
product expansion is on the same footing as that of lifetimes.

\subsection{Some differences between the neutral meson systems}

The general solution for the eigenvalues is~\cite{CPV-TheBook}
\beqa\label{mixparam}
(\Delta m)^2 - {(\Delta\Gamma)^2\over 4} 
  &=& 4\, |M_{12}|^2 - |\Gamma_{12}|^2\,, \nn\\*
\Delta m\, \Delta\Gamma &=& 4\, {\rm Re} (M_{12} \Gamma_{12}^*)\,, \nn\\*
{q^2\over p^2} &=& \frac{2 M_{12}^*-i\,\Gamma_{12}^*}{2M_{12}-i\,\Gamma_{12}}\,.
\eeqa
The behavior of these solutions is different depending on the magnitudes of
$\Delta m$ and $\Delta\Gamma$.  The mixing parameters satisfy $|\Delta\Gamma|
\ll \Delta m$ for $B_{d,s}$ mixing, $\Delta\Gamma \approx -2\, \Delta m$ for $K$
mixing, and the current data is not yet conclusive for $D$ mixing.

In the $B_{d,s}$ systems $\Delta m \gg |\Delta\Gamma|$ both in the SM and
beyond.  The first two relations in Eq.~(\ref{mixparam}) imply that this is
equivalent to $|\Gamma_{12} / M_{12}| \ll 1$.  In this case,
\beqa
\Delta m &=& 2\, |M_{12}|\, (1 + \ldots)\,, \nn\\*
\Delta\Gamma &=& -2\, |\Gamma_{12}|\, \cos\phi_{12}\, (1 + \ldots) \,,
\eeqa
where the ellipses denote terms suppressed by powers of $\Gamma_{12} / M_{12}$. 
In $B_{d,s}$ mixing $\phi_{12}$ is suppressed by $m_c^2/m_b^2$, and in addition
by $|V_{us}/V_{ud}|^2$ for $B_s$.  Thus, NP in $\phi_{12}$ can only suppress
$|\Delta\Gamma_{B_s}|$~\cite{Grossman:1996er}.  Moreover,
\beq
{q^2\over p^2} = {(M_{12}^*)^2 \over |M_{12}|^2}\, (1 + \ldots)\,,
\eeq
so time dependent $CP$ asymmetry measurements have good sensitivity to NP in
$M_{12}$, e.g., $\arg \lambda_{\psi K} \propto \phi_{12}$.

If $\Delta m \ll |\Delta\Gamma|$ holds, the solution would be rather different. 
The first two relations in Eq.~(\ref{mixparam}) imply that this is equivalent to
$|M_{12}/\Gamma_{12}| \ll 1$.  In this case~\cite{Bergmann:2000id}
\beqa
\Delta m &=& 2\, |M_{12}|\, |\cos\phi_{12}|\, (1 + \ldots)\,, \nn\\*
\Delta\Gamma &=& -2\, |\Gamma_{12}|\; {\rm sgn}(\cos\phi_{12})\, (1 + \ldots)\,,
\eeqa
where the ellipses denote terms suppressed by powers of $M_{12}/\Gamma_{12}$.
The signs are chosen to ensure $\Delta m > 0$.  Moreover,
\beq
{q^2\over p^2} = {(\Gamma_{12}^*)^2\over |\Gamma_{12}|^2}\, (1 + \ldots)\,,
\eeq
so $q/p$ depends only weakly on $M_{12}$.  Neglecting $CP$ violation in $D$
decay, $\Delta m \ll |\Delta\Gamma|$ would imply, e.g.,
\beq
\arg \lambda_{K^+K^-} \propto 2\, \bigg|{M_{12}\over
  \Gamma_{12}}\bigg|^2 \sin(2\phi_{12})\,.
\eeq
We learn that if $|\Delta\Gamma| \gg \Delta m$ then the sensitivity to NP in
$M_{12}$ is suppressed by $\Delta m / \Delta\Gamma$ even if NP dominates
$M_{12}$~\cite{Bergmann:2000id}.  We also learn that $\Delta m \gg
|\Delta\Gamma|$ or $\Delta m \ll |\Delta\Gamma|$ necessarily imply $|q/p|
\approx 1$, while if $|\Delta\Gamma| \sim \Delta m$ then $|q/p|$ may be far from
$1$ and large $CP$ violating effects in mixing are possible in principle.

The present data imply $|\Delta\Gamma/(2\Gamma)| \sim 0.01$ at $3.5\sigma$ in
the $D^0$ system, while the indication for $\Delta m \neq 0$ is about $2\sigma$,
so the values of $\Delta\Gamma$ and $\Delta m$ are not yet settled.  Thus,
instead of $|D_{L,H}\rangle$, we label the states as $|D_{1,2}\rangle = p\,
|D^0\rangle \pm q\, |\D0bar\rangle$.  The fact that $\Delta\Gamma/\Delta m$
affects significantly the sensitivity of any observable to a possible $CP$
violating NP contribution in $M_{12}$ provides a strong reason to pin down
$\Delta m$ and $\Delta\Gamma$.

\section{\boldmath $D^0$\,--\,$\D0bar$ mixing: measurements and their
interpretations}

The dimensionless mass and width difference parameters that characterize
$D^0$\,--\,$\D0bar$ mixing are
\beq\label{xydef}
x = {\Delta m \over \Gamma}\,, \qquad
  y = {\Delta \Gamma \over 2\Gamma}\,,
\eeq
and it has been often stated that $x$ and $y$ are expected to be well below
$10^{-2}$ in the SM.

The $D^0$ meson system is unique among the neutral mesons in that it is the only
one in which mixing proceeds via intermediate states with down-type quarks (or
up-type squarks in supersymmetric models). The mixing is very slow in the SM,
because the third generation plays a negligible role in FCNC box and penguin
diagrams due to the smallness of $|V_{ub} V_{cb}| = {\cal O}(10^{-4})$, so the
GIM cancellation is very effective.  In the SM, $x$ and $y$ have two powers of
Cabibbo suppression and only arise at second order in $SU(3)$
breaking~\cite{Falk:2001hx},
\beq\label{xygen}
x,\, y \,\sim\, \sin^2\theta_C \times [SU(3) \mbox{ breaking}]^2\,,
\eeq
where $\theta_C$ is the Cabibbo angle.  The theoretical predictions have large
uncertainties and depend crucially on estimating the size of $SU(3)$ breaking. 
Possible NP in $D^0$\,--\,$\D0bar$ mixing can modify $M_{12}$, but its effect on
$\Gamma_{12}$ is generically suppressed by an additional loop (penguin vs.\ tree
decay).   (See Ref.~\cite{Golowich:2006gq} for more discussion.)  Thus, at the
current level of sensitivity, $\Delta m \gg \Delta\Gamma$ would indicate NP,
while $\Delta\Gamma \gsim \Delta m$ would signal large SM contributions.  As
explained above, although $y$ is expected to be determined by SM processes, the
ratio $y/x$ significantly affects the sensitivity of mixing to new physics.

To study various observables that involve mixing and decay, it is convenient to
expand the time dependence of the decay rates in the small parameters, $x$ and
$y$.  Throughout this talk we neglect $CP$ violation in $D$ decays (direct $CP$
violation), unless explicitly stated otherwise.  Then we can 
write~\cite{Bergmann:2000id}
\beqa\label{lambdas}
\lambda_{K^-\pi^+} &=& \sqrt R\, |q/p|\, e^{-i(\delta-\phi)} , \nn\\
\lambda_{K^+\pi^-}^{-1} &=& \sqrt R\, |p/q|\, e^{-i(\delta+\phi)} , \nn\\
\lambda_{K^+ K^-} &=& -\, |q/p|\, e^{i\phi} .
\eeqa
For doubly-Cabibbo-suppressed (DCS) decays (i.e., $c\to d\bar s u$ or mixing
followed by $\bar c \to \bar s d\bar u$), we can expand in 
$|\lambda_{K^-\pi^+}|$ and $|\lambda_{K^+\pi^-}^{-1}|$, which are ${\cal
O}(\tan^2\theta_C)$,
\begin{widetext}\vspace*{-24pt}
\beqa\label{DCS1}
\Gamma\big[D^0(t) \to K^+\pi^-\big] &=& e^{-\Gamma t}\, |\bar A_{K^+\pi^-}|^2\,
  \bigg|{q\over p}\bigg|^2\, \bigg\{ |\lambda_{K^+\pi^-}^{-1}|^2 
  + \big[{\rm Re}(\lambda_{K^+\pi^-}^{-1})\, y 
  + {\rm Im}(\lambda_{K^+\pi^-}^{-1})\, x \big]\, \Gamma\, t
  + {y^2+x^2\over 4}\, (\Gamma\, t)^2 \bigg\} \nn\\*
&=& e^{-\Gamma t}\, |A_{K^-\pi^+}|^2
  \bigg[ R + \sqrt R\, \bigg|{q\over p}\bigg|\, 
  (y' \cos\phi - x' \sin\phi)\, \Gamma\, t
  + \bigg|{q\over p}\bigg|^2\, {y^2+x^2\over 4}\, (\Gamma\, t)^2\bigg] , \\
\Gamma\big[\D0bar(t) \to K^-\pi^+\big] &=& e^{-\Gamma t}\, |A_{K^-\pi^+}|^2\,
  \bigg|{p\over q}\bigg|^2\, \bigg\{ |\lambda_{K^-\pi^+}|^2 
  + \big[{\rm Re}(\lambda_{K^-\pi^+})\, y 
  + {\rm Im}(\lambda_{K^-\pi^+})\, x \big]\, \Gamma\, t
  + {y^2+x^2\over 4}\, (\Gamma\, t)^2 \bigg\} \nn\\*
&=& e^{-\Gamma t}\, |A_{K^-\pi^+}|^2
  \bigg[ R + \sqrt R\, \bigg|{p\over q}\bigg|\, 
  (y' \cos\phi + x' \sin\phi)\, \Gamma\, t 
  + \bigg|{p\over q}\bigg|^2\, {y^2+x^2\over 4}\, (\Gamma\, t)^2 \bigg] .
\label{DCS2}
\eeqa
Here $x' = x \cos\delta + y \sin\delta$, $y' = y \cos\delta - x \sin\delta$,
and $\delta = -\arg \big(\lambda_{K^-\pi^+}\, \lambda_{K^+\pi^-}^{-1}\big)/2$ is
the strong phase between the Cabibbo-favored (CF) and the DCS amplitudes.  The
first terms on the right-hand sides come from the direct DCS decay, the last
terms from mixing followed by CF decay, and the middle ones from their
interference.  For singly-Cabibbo-suppressed (SCS) decays (e.g., $c\to s\bar s
u$ or mixing followed by $\bar c \to \bar s s\bar u$), the rates are
\beqa\label{SCS1}
\Gamma\big[D^0(t) \to K^+K^-\big] &=& e^{-\Gamma t}\, |A_{K^+ K^-}|^2\,
  \bigg\{ 1 + \big[{\rm Re}(\lambda_{K^+ K^-})\, y 
  - {\rm Im}(\lambda_{K^+ K^-})\, x \big]\, \Gamma\, t \bigg\} \nn\\*
&=& e^{-\Gamma t}\, |A_{K^+ K^-}|^2\,
  \bigg[ 1 - \bigg|{q\over p}\bigg|\, (y\cos\phi-x\sin\phi)\, \Gamma\,t\bigg],\\
\Gamma\big[\D0bar(t) \to K^+K^-\big] &=& e^{-\Gamma t}\, |\bar A_{K^+ K^-}|^2\,
  \bigg\{ 1 + \big[{\rm Re}(\lambda_{K^+ K^-}^{-1})\, y 
  - {\rm Im}(\lambda_{K^+ K^-}^{-1})\, x \big]\, \Gamma\, t \bigg\} \nn\\*
&=& e^{-\Gamma t}\, |A_{K^+ K^-}|^2\,
  \bigg[1 - \bigg|{p\over q}\bigg|\, (y\cos\phi+x\sin\phi)\, \Gamma\,t\bigg].
\label{SCS2}
\eeqa
\end{widetext}
Finally, for Cabibbo-favored (CF) decays ($c\to s\bar d u$),
\beq\label{CF}
\Gamma\big[D^0(t)\to K^-\pi^+\big]
  = \Gamma\big[\D0bar(t)\to K^+\pi^-\big] \propto e^{-\Gamma t} . 
\eeq

The first lines in Eqs.~(\ref{DCS1}) -- (\ref{SCS2}) [but not
Eq.~(\ref{lambdas})] are valid if there is direct $CP$ violation.  In the limit
of $CP$ conservation, choosing $\phi=0$ in Eqs.~(\ref{lambdas}) -- (\ref{CF})
amounts to defining $|D_1\rangle = CP$-odd and $|D_2\rangle = CP$-even, since
$\langle K^+ K^- | {\cal H} | D_{1,2}\rangle = p\, A_{K^+ K^-} (1 \pm
\lambda_{K^+ K^-})$, while $\phi=\pi$ is the opposite choice.

\subsection{Mixing parameters from lifetimes}
\label{sec:yCP}

Several experiments measured the $D$ meson lifetime, $\hat\tau(D\to f)$, by
fitting single exponential time dependences to the decay rates to $CP$
eigenstates and flavor specific modes.  Two important observables are
\beqa\label{yCP}
y_{CP} &=& {\hat \tau(D \to \pi^+ K^-) \over \hat\tau(D \to K^+ K^-)} - 1 \\*
&=& {y\cos\phi\over2} \bigg(\bigg|{q\over p}\bigg|+\bigg|{p\over q}\bigg|\bigg)
  - {x\sin\phi\over2} \bigg(\bigg|{q\over p}\bigg|-\bigg|{p\over q}\bigg|\bigg)
  \,,\nn\\
A_\Gamma &=& {\hat\tau(\D0bar \to K^+ K^-) - \hat\tau(D^0 \to K^+ K^-)
  \over \hat\tau(\D0bar \to K^+ K^-) + \hat\tau(D^0 \to K^+ K^-) } \\*
&=& {y\cos\phi\over2} \bigg(\bigg|{q\over p}\bigg|-\bigg|{p\over q}\bigg|\bigg)
  - {x\sin\phi\over2} \bigg(\bigg|{q\over p}\bigg|+\bigg|{p\over q}\bigg|\bigg)
  \,.\nn \label{AG}
\eeqa
Here $y_{CP}$ is related to the lifetime difference of the (approximately)
$CP$-odd and even $D$ states.  If $CP$ is conserved, $A_\Gamma=0$ and $y_{CP} =
\pm\, y$ (depending on whether $\phi$ is 0 or $\pi$).  The current data,
\beqa\label{yCPAGres}
y_{CP} &=& 0.0112 \pm 0.0032 \quad \mbox{\cite{hfagc}}\,, \nn\\
A_\Gamma &=& 0.0001 \pm 0.0034 \quad \mbox{\cite{Staric:2007dt}}\,,
\eeqa
show $y_{CP} \neq 0$ at the $3.5\sigma$ level.  The quoted value of 
$y_{CP}$ is the average of the Belle, BaBar, CLEO, FOCUS, and E791
measurements~\cite{Staric:2007dt,yCPold}.

Given that $y_{CP} \neq 0$ and $A_\Gamma$ is consistent with 0, it is suggestive
(though not yet conclusive) that $y \sim 0.01$ and $CP$ violation in mixing
or/and $x$ are small.

\subsection{\boldmath Mixing parameters from $D\to K^+\pi^-$}

\begin{figure}[tb]
\centering
\includegraphics*[width=.9\columnwidth]{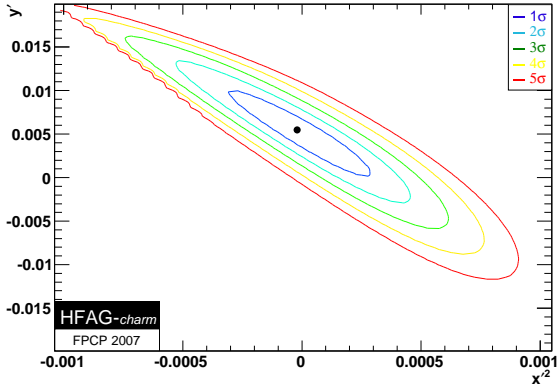}
\caption{Confidence level of the average of the BaBar and Belle measurements
of $x^{\prime2}$ and $y'$ (from~\cite{hfagc}).}
\label{fig:Kpi}
\end{figure}

One can also measure the time dependence of doubly-Cabibbo-suppressed decays,
such as $D^0\to K^+ \pi^-$.  In the $CP$ conserving limit, the measurements are
sensitive to $y'$, $x^{\prime2}$, and $R$ (recall $x^{\prime2}+y^{\prime2} =
x^2+y^2$).  The most significant measurement to date from
BaBar~\cite{Aubert:2007wf}
\beq
y' = 0.0097\pm 0.0054 , \quad
  x'{}^2 = (-2.2 \pm 3.7)\times 10^{-4} ,
\eeq
gives $3.9\sigma$ evidence for mixing, due to the strong correlation between
$x^{\prime2}$ and $y'$.  To illustrate this, Fig.~\ref{fig:Kpi} shows the confidence
level of $x^{\prime2}$ and $y'$ combining the two most sensitive
measurements~\cite{Aubert:2007wf,Zhang:2006dp}, giving an over $4\sigma$
deviation from the no-mixing hypothesis.

If $\sin\phi \neq 0$ then the measurements have linear sensitivity to both $x'$
and $y'$.  By virtue of Eqs.~(\ref{DCS1}) and (\ref{DCS2}), allowing $CP$
violation in mixing increases the number of fit parameters from 3 to 5 (adding
$\phi$ and $|q/p|$). Equivalently, the experimental analyses fit for
\beqa\label{xyppm}
x^{\prime\pm} &=& \big|{q / p}\big|^{\pm1} (x'\cos\phi \pm y'\sin\phi)\,,
  \nn\\*
y^{\prime\pm} &=& \big|{q / p}\big|^{\pm1} (y'\cos\phi \mp x'\sin\phi)\,,
\eeqa
and find consistent results with those in Fig.~\ref{fig:Kpi} and no hint of $CP$
violation.  Note that the experimental
papers~\cite{Aubert:2007wf,Zhang:2006dp,Godang:1999yd} use 6-parameter fits,
including two parameters, $R_D$ and $A_D$, instead of $R$.  Unless there is $CP$
violation in $D$ decay, $R_D = R$ and $A_D = 0$, so it would be very interesting
to know the results of the 5-parameter fits with $A_D = 0$ enforced.  (This may
be similar to the early $B\to J/\psi K_S$ analyses, when $S_{\psi K}$ was
measured both with and without imposing $|\lambda_{\psi K}|=1$.  It would be
interesting to see if imposing $A_D = 0$ would have a significant impact.)

\subsection{\boldmath Mixing parameters from the $D\to K_S\, \pi^+\pi^-$
Dalitz analysis}

Similar to the measurement of the CKM angle $\gamma$ from $B^\pm\to
D(K_S\pi^+\pi^-) K^\pm$, one can also search for $D^0$\,--\,$\D0bar$ mixing in
the same $D$ decay.  The Dalitz plot analysis is based on writing the amplitudes
as
\beqa
&&\hspace*{-.2cm} \langle K_S\pi^+\pi^- |\, {\cal H}\, | D_0(t) \rangle \\ 
&&{} = A_{K_S\pi^+\pi^-}
  \big[e^{-(\Gamma_1/2 + im_1)t} + e^{-(\Gamma_2/2 + im_2)t}\big] /2 \nn\\
&&{} + (q/p)\, \bar A_{K_S\pi^+\pi^-}
  \big[e^{-(\Gamma_1/2 + im_1)t} - e^{-(\Gamma_2/2 + im_2)t}\big] /2 , \nn
\eeqa
and similarly for $\D0bar(t)$.  Denoting $m_\pm = m_{K_S\pi^\pm}$, with no
direct $CP$ violation, $A(m_+, m_-) = \bar A(m_-, m_+)$.  The amplitude is
modelled by a sum of resonances, $\sum_j a_j e^{i\delta_j} {\cal A}^j$, where
${\cal A}^j$ is the model for each resonance that depends on $m_+$ and $m_-$,
while $a_j$ and $\delta_j$ are its amplitude and strong phase.  Thus, the rate
depends on interferences involving rapidly varying known strong phases related
to the resonances (i.e., $\Gamma_{K^*} \ll m_D$), and is sensitive to $x$ and
$y$, including the sign of $x/y$.\footnote{Recall that to measure ${\rm
sgn}(m_{K_L} - m_{K_S})$, input on phase shifts had to be used~\cite{KLKS}, and
it was only determined in 1966, even after the discovery of $CP$ violation.} 
With the time dependence of rates to $CP$ eigenstates (e.g., $\rho^0 K_S$), all
signs can be resolved (except the unphysical $\{x,\, y,\, q/p\}\to \{-x,\, -y,\,
-q/p\}$).

The analysis relies on the amplitude throughout the Dalitz plot, but its
modelling has only been tested with the rates so far.  In the region of the
Dalitz plot corresponding to large $K^{**}$ masses ($K^{**}$ denotes heavy kaon
states which decay to $K_S\pi$) the ratio of the DCS and CF rates is
significantly enhanced in the Belle model~\cite{Abe:2007rd} compared to that for
$D\to K\pi$.\footnote{I thank Bostjan Golob for drawing my attention to this.} 
While this is possible theoretically, it is less pronounced in the BaBar
model~\cite{Poluektov:2006ia}.  Data on $CP$-tagged $D\to K_S\pi^+\pi^-$ decays
expected soon from CLEO-c could help reduce the uncertainties.  (With more data,
one may also attempt a model independent analysis, as for the extraction of the
CKM angle $\gamma$~\cite{Giri:2003ty}.)

The first significant result is from Belle~\cite{Abe:2007rd}
\beq\label{xyres}
x = 0.0080 \pm 0.0034\,, \quad y = 0.0033 \pm 0.0028\,,
\eeq
which is $2.7\sigma$ from the no-mixing hypothesis.  The 95\%\,CL intervals are
$0<x<0.016$ and $-0.0035<y<0.010$. The preliminary result allowing for all but
direct $CP$ violation (the analog of the 5-parameter fit for $K\pi$ with $A_D =
0$, advocated above) is consistent with this result, and yields~\cite{alan}
\beqa\label{xyqpres}
x &=& 0.0081 \pm 0.0034, \qquad  |q/p| = 0.95^{+0.22}_{-0.20}\,, \\*
y &=& 0.0037 \pm 0.0029, \quad   \arg(q/p) = -0.03 \pm 0.19, \nn
\eeqa
where $\arg(q/p)$ is to be understood in the phase convention in which $CP\,
|D^0\rangle = |\D0bar\rangle$.  This shows no hint of $CP$ violation yet.

\subsection{Other measurements and some interpretation}

Several other measurements are sensitive to $D^0$\,--\,$\D0bar$ mixing.  The
``wrong sign" semileptonic $D^0$ rate (the phenomenon by which $B^0$ mixing was
discovered) has only quadratic sensitivity to $x$ and $y$, giving $x^2+y^2 =
(3.5 \pm 7.7) \times 10^{-4}$~\cite{hfagc}.  In the limit of very large data
sets, measurements with linear sensitivity are expected to give the best
constraints.

Other Dalitz analyses, such as $D^0\to K^+\pi^-\pi^0$~\cite{Brandenburg:2001ze}
and measurements of $D^0\to K^+\pi^-\pi^+\pi^-$ may also prove useful in pinning
down the mixing parameters by providing complementary information to the
measurements discussed above.

Combining all experimental results obtained without allowing for $CP$ violation,
HFAG finds a $5.7\sigma$ signal for $D^0$\,--\,$\D0bar$ mixing, with the
projections~\cite{hfagc}
\beq\label{hfagfit}
x = 0.0087^{+0.0030}_{-0.0034}\,, \quad y = 0.0066 \pm 0.0021\,.
\eeq
As the experimental uncertainties decrease, it will be interesting to allow for
$CP$ violation in mixing (i.e., $|q/p|\neq 1$ and $\sin\phi \neq 0$) in the
fits. If the $y$ term dominates $y_{CP}$ in Eq.~(\ref{yCP}) and $y'$ dominates
$y^{\prime\pm}$ in Eq.~(\ref{xyppm}) then the $CP$ violating
ratios~\cite{Nir:2007ac}
\beqa
{A_\Gamma\over y_{CP}} &\approx&
  {|q/p|^2-1\over |q/p|^2+1} - {x\over y} \tan\phi\,, \nn\\
{y^{\prime+}-y^{\prime-}\over y^{\prime+}+y^{\prime-}}&\approx&
  {|q/p|^2-1\over |q/p|^2+1} - {x'\over y'} \tan\phi\,,
\eeqa
give simple constraints on $|q/p|$ and $\phi$.  This would of course be taken
into account in a fit that allows $CP$ violation and includes all correlations
between the measurements.   While the fit assuming no $CP$ violation giving
Eq.~(\ref{hfagfit}) has a good $\chi^2$, I would caution about over-interpreting
it until we see how the difference between $y_{CP}$ in Eq.~(\ref{yCPAGres}) and
$y$ in Eq.~(\ref{xyres}) will change as the uncertainties decrease.

Given that the measured values of the $D^0$\,--\,$\D0bar$ mixing parameters may
be due to long distance hadronic physics, to set constraints on new
physics~\cite{Ciuchini:2007cw}, one has to assume that there is no cancellation
between the NP and the SM contributions, and can only demand that the NP
contribution does not exceed the measured values.  This situation could change
when $\Delta\Gamma$ and $\Delta m$ become better known, and especially if $CP$
violation is observed.  Thus, it will be very interesting to robustly establish
the values of the mixing parameters as more experimental results appear.

\subsection{\boldmath Calculations of $\Delta\Gamma_D$ and $\Delta m_D$}

The reason it is notoriously hard to calculate $x$ and $y$ in the SM is that the
charm quark is neither heavy nor light enough to trust the theoretical tools
applicable in these two limits. The lowest order short-distance calculation of
the box diagram gives tiny results,
\beq
x_{\rm box} \propto {m_s^2 \over m_W^2} \times {m_s^2 \over m_c^2}\,,
  \qquad  y_{\rm box} \propto {m_s^2 \over m_c^2}\, x_{\rm box}\,,
\eeq
yielding few\,$\times 10^{-5}$ and few\,$\times 10^{-7}$, respectively.  The
$m_s^4$ suppression of $x_{\rm box}$ arises, because at short distances, above
the chiral symmetry breaking scale, each power of $SU(3)$ breaking ($U$-spin
breaking) required by Eq.~(\ref{xygen}) is proportional to $m_s^2$ instead of
$m_s$~\cite{Georgi:1992as}. An additional $m_s^2$ suppression of $y_{\rm
box}/x_{\rm box}$ arises from the helicity suppression of the decay of a scalar
meson into a massless fermion pair; this is why at leading order in the OPE,
$y_{\rm box} \ll x_{\rm box}$.

\begin{table}\vspace*{-8pt}
\caption{Enhancement of $\Delta m$ and $\Delta\Gamma$ relative to the box
diagram (4-quark operator) at higher orders in the OPE ($\Lambda$ is a 
hadronic scale around 1GeV and $\beta_0=11-2n_f/3=9$).}
\label{opetable}
\tabcolsep 6pt
\begin{tabular}{c|ccc} \hline\hline
Ratio  &  ~4-quark~  &  6-quark  &  8-quark \\ \hline\hline
\raisebox{0pt}[15pt][8pt]{$\ds{\Delta m\over \Delta m_{\rm box}}$}
  &  1  &  $\ds{\Lambda^2\over m_s m_c}$
  &  $\ds{\alpha_s\over4\pi}\, \Big({\Lambda^2\over m_s m_c}\Big)^2$ \\ \hline
\raisebox{0pt}[15pt][8pt]{$\ds {\Delta\Gamma\over \Delta m}$}
  &  $\ds{m_s^2\over m_c^2}$
  &  $\ds{\alpha_s\over4\pi}$  &  $\ds\beta_0\,{\alpha_s\over 4\pi}$ \\ 
  \hline\hline
\end{tabular}\vspace*{-3pt}
\end{table}

It was recognized by Georgi that higher order contributions to $x$ and $y$ in
the OPE have fewer powers of $m_s$ suppressions, since the chiral suppressions
can be lifted by quark condensates instead of mass
insertions~\cite{Georgi:1992as}.  The parametric enhancement of the subleading
terms are summarized in Table~\ref{opetable}~\cite{Falk:2001hx}, which shows
that the 8-quark operator contributions to $x$ and $y$ are only suppressed by
$m_s^2$, the minimal possible power.  Thus,  these higher dimension operators
give the dominant contributions.  Using naive dimensional analysis ($\Lambda
\sim 4\pi f_\pi$) and different assumptions to estimate the matrix elements, one
can find smaller~\cite{Ohl:1992sr} or larger enhancements~\cite{Bigi:2000wn},
yielding up to
\beq 
x \sim y \sim 10^{-3}\,. 
\eeq 
Since there are several unknown matrix elements which are hard to estimate,
these results are at best useful to understand the orders of magnitudes of $x$
and $y$, but not for obtaining reliable SM predictions (even at the factor of
2--3 level).

In a long-distance analysis, instead of assuming that the $D$ meson is heavy
enough for duality to hold between the partonic rate and the sum over hadronic
final states, one examines certain exclusive decay modes.  There is a
contribution to $y$ from all final states common to $D^0$ and $\D0bar$ decay,
\beq\label{yexcl}
y = {\sum_n \rho_n \langle\D0bar|\, {\cal H}_w|n\rangle
  \langle n|\, {\cal H}_w|D^0\rangle \over \sum_n \Gamma(D^0\to n)}\,,
\eeq
where $\rho_n$ is the phase space available to the state $n$ (we neglect $CP$
violation, and choose $\Gamma_{12}$ to be real).  We denote by $y_{F,R}$ the
expression in Eq.~(\ref{yexcl}) with the sum over $n$ restricted to states $F$
(e.g., certain number of pseudoscalar or vector mesons) in the $SU(3)$
representation $R$, $n \in F_R$.  The $y_{F,R}$ are the ``would-be" values of
$y$, if $D$ only decayed to $F_R$.  In the $SU(3)$ limit, $y_{F,R}=0$.  Since
$D$ decays are not dominated by a few final states and there are cancellations
between states within a given $SU(3)$ multiplet, to make a reliable estimate one
would need to know the contributions of many states with high precision.  In the
absence of sufficiently precise data on the rates and strong phases, one has to
use assumptions.

The importance of $SU(3)$ cancellations in the magnitudes and phases of matrix
elements can be illustrated by $D$ decays to a pair of charged $\pi$'s and
$K$'s.  The $SU(3)$ breaking is very large in ${\cal B}(D^0 \to K^+ K^-) / {\cal
B}(D^0 \to \pi^+ \pi^-) \approx 2.8$, which is unity in the $SU(3)$
limit.\footnote{The $SU(3)$ breaking in the matrix elements may actually be
modest, although this ratio is far from the $SU(3)$
limit~\cite{Savage:1991wu}.}  This was the basis for the claim that $SU(3)$ is
not applicable to $D$ decays, so $x,y \sim 10^{-2}$ is
possible~\cite{Wolfenstein:1985ft}.  (However, as we show below, these states
alone are unlikely to give so large $x$ and $y$, due to their small rates.)  The
value of $y$ corresponding to decays to $\pi^+\pi^-$, $\pi^\pm K^\mp$, and $K^+
K^-$ is
\beqa
y_{\pi K} &\propto& {\cal B}(D^0 \to \pi^+\pi^-) + {\cal B}(D^0 \to K^+K^-) \\*
&-& 2\cos\delta\, \sqrt{{\cal B}(D^0 \to K^-\pi^+)\, 
  {\cal B}(D^0 \to K^+\pi^-)} \,, \nn
\eeqa
where $\delta$ is the strong phase between the CF and DCS amplitudes defined
after Eq.~(\ref{DCS2}), which vanishes in the $SU(3)$ limit.  The experimental
central values~\cite{pdg} yield $(5.2 - 4.7 \cos\delta) \times 10^{-3}$.  For
small $\delta$, there is a significant cancellation, and the result is
consistent with zero within $1\sigma$, even though the individual rates badly
violate $SU(3)$.  One cannot use, however, this exclusive approach to reliably
predict $x$ or $y$, since the estimates are very sensitive to $SU(3)$ breaking
in poorly known strong phases and DCS rates.

The cancellations that give $y_{F,R}=0$ in the $SU(3)$ limit depend on both the
matrix elements and the phase space, $\rho_n$, in Eq.~(\ref{yexcl}).  We cannot
estimate model independently the $SU(3)$ violation in matrix elements, but that
in the phase space is calculable, as it mainly depends on the hadron masses in
the final states, and can be computed with mild assumptions about the momentum
dependence of the matrix elements.  Incorporating the true values of $\rho_n$ in
Eq.~(\ref{yexcl}) is a calculable source of $SU(3)$ breaking.\footnote{Such
phase space differences can explain the large $SU(3)$ breaking between the
measured $D\to K^*\ell\bar\nu$ and $D\to \rho\ell\bar\nu$ rates, assuming no
$SU(3)$ breaking in the form factors~\cite{Ligeti:1997aq}.  The lifetime ratio,
$\tau_{D_s} / \tau_{D^0}$, may also be explained this
way~\cite{Nussinov:2001zc}.}  This contribution to $y$ due to $SU(3)$ violation
in phase space is negligible for two-body pseudoscalar final states, but can be
of the order of a percent for final states with masses near $m_D$.

To illustrate some aspects of this analysis~\cite{Falk:2001hx,Ligeti:2002gc}, consider the above example of
the $U$-spin doublet of charged kaons and pions,
\beqa
y_{\pi K} &=& \sin^2\theta_C\, \big[ \Phi(\pi^+,\pi^-) + \Phi(K^+,K^-) \nn\\*
&&{} - 2\Phi(K^+,\pi^-) \big] \,\big/\, \Phi(K^+,\pi^-) \,,
\eeqa
where $\Phi$ is the phase space. This model sets $\delta=0$, so it gives $y_{\pi
K} \sim - 0.01 \sin^2\theta_C$, a tiny result.  For representations in which
some states are not allowed by phase space, $SU(3)$ breaking is large.  For
example, for 4 pseudoscalar mesons the phase space depends very strongly on the
number of kaons and vanishes for $D\to 4K$ ($m_{4K} > m_D$), giving $y_{4P} =
{\cal O}(\sin^2\theta_C)$.  Clearly, this enhancement of $y$ is a ``threshold
effect", which would be small if $m_c$ were heavier, but is significant for the
physical value of $m_c$.  Not all final states which may give large
contributions were considered in Ref.~\cite{Falk:2001hx}; e.g., ${\cal
B}(D^0\to  K^- a_1^+) = (7.5 \pm 1.1)\%$, although its phase space is very
small.  Since 4 pseudoscalars account for $\sim$10\% of the $D$ width, the
contribution of these states alone to $y$ can be near 0.01.

Thus, we conclude that $y \sim 0.01$ is natural in the SM.  An order of
magnitude smaller result would require significant cancellations, which would
only be expected if they were enforced by the OPE.  

To connect the calculation of $y$ to $x$, a dispersion relation can be proven in
HQET, which relates $\Delta m$ to an integral of $\Delta\Gamma$ over the mass
$M$ of a heavy ``would-be $D$ meson"~\cite{Falk:2004wg}
\beq
\Delta m = -\frac{1}{2\pi}\, {\rm P}\! \int_{2m_\pi}^\infty \d M\,
  \frac{\Delta\Gamma(M)}{M-m_D} + \ldots \,.
\eeq
Modelling that phase space is the only source of $SU(3)$ breaking, the
calculation of $x$ based on this relation is more model dependent than that of
$y$.  Unlike the estimate of $y$, the hadronic matrix elements do not cancel
in $x$, since some assumptions about the $M$-dependence of the rates has to be
made.  The most significant (tractable) contributions come again from 4-body
final states, which can give $x$ comparable in magnitude to $y$ (thought
typically $0.1 < x/y \lsim 1$)~\cite{Falk:2004wg}.

\subsection{\boldmath Summary for $D^0$\,--\,$\D0bar$ mixing}

\begin{itemize}\itemsep 0pt

\item The central values of recent experimental results may be due to SM
physics.

\item It is possible that $\Delta\Gamma/(2\Gamma) \sim 0.01$ in the SM (some
calculable contributions are of this size).

\item It is likely that $\Delta m \lsim \Delta\Gamma$ in the SM (though this
relies on significant assumptions).

\item If $x<y$ then sensitivity to NP is reduced, even if NP dominates $M_{12}$.

\item The SM predictions of $\Delta m$ and $\Delta\Gamma$ remain uncertain, so
their measurements alone (especially if $\Delta m \lsim \Delta\Gamma$) cannot be
interpreted as NP.

\item It is important to improve the constraints on both $\Delta\Gamma$ and
$\Delta m$, and to look for $CP$ violation, which remains a potentially robust
signal of NP.

\end{itemize}\vspace*{-6pt}

\section{\boldmath $B_s^0$\,--\,$\B0bar_s$ mixing}

The $B_s^0$ and $\B0bar_s$ mesons oscillate about 25 times before they decay,
which made measuring the oscillation frequency very challenging.  The
measurement~\cite{Abulencia:2006ze}
\beq\label{dms}
\Delta m_s = (17.77 \pm 0.10 \pm0.07)\,{\rm ps}^{-1}\,,
\eeq
is a key to test and overconstrain the CKM matrix and the SM description of $CP$
violation.  Amusingly, the experimental uncertainty $\sigma(\Delta m_s) = 0.7\%$
is already smaller than $\sigma(\Delta m_d) = 0.8\%$, which has been measured
for over 20 years.

To interpret the result in Eq.~(\ref{dms}) in terms of CKM parameters, the
largest uncertainty comes from the hadronic matrix element $f_{B_s} \sqrt{B_s}$,
whose error is around 15\%.  To reduce this (and because in the context of
testing the SM one is more interested in the value of $|V_{td} V_{tb}|$ than
$|V_{ts} V_{tb}|$), one considers the ratio $\Delta m_s/\Delta m_d$, which is
precisely calculable in terms of $|V_{td}/V_{ts}|$ and $\xi = (f_{B_s}
\sqrt{B_s}) / (f_{B_d} \sqrt{B_d})$.  Here $\xi$ quantifies $SU(3)$-breaking
corrections to the ratio of matrix elements, which can be calculated more
accurately in lattice QCD (LQCD) than the matrix elements separately (the
calculation of chiral logs predicts $\xi \sim 1.2$~\cite{Bslogs}).  CDF infers
from its measurement of $\Delta m_s$ the ratio of CKM elements,
\beq
|V_{td}/V_{ts}| = 0.206\pm0.001\mbox{(exp)}\,^{+0.008}_{-0.006}\mbox{(theo)}\,,
\eeq
where the error is dominated by the theoretical uncertainty of $\xi =
1.21^{+0.047}_{-0.035}$~\cite{Okamoto:2005zg}, used by CDF. The CDF, D\O, ALEPH,
and DELPHI experiments have also measured the $B_s^0$ lifetimes in $CP$-even,
$CP$-odd, and flavor specific final states, yielding~\cite{hfag}
\beq
\Delta\Gamma_s^{CP} = (0.071^{+0.053}_{-0.057})\, {\rm ps}^{-1} ,
\eeq
where $\Delta\Gamma_s^{CP} = \Gamma_{CP+} -  \Gamma_{CP-} = -\cos\phi_{12}\,
\Delta\Gamma_s$~\cite{Grossman:1996er,Dunietz:2000cr}.  This is similar to the
measurement of $y_{CP}$ in Sec.~\ref{sec:yCP}.

The mixing in the $B_d$ and $B_s$ systems are short distance dominated, so the
theory errors in interpreting $\Delta m_{d,s}$ are suppressed compared to the
measured values.  (This is in contrast with $\Delta m_D$ and
$\epsilon'/\epsilon$, where due to hadronic uncertainties we only know at
present that the NP contributions do not exceed the observations.)  The
interpretation of the measurement of $\Delta\Gamma_s^{CP}$ (or $\Delta\Gamma_s$)
relies on the calculation of $\Gamma_{12}$, which is on the same footing as that
of heavy hadron lifetimes.  This makes it important to resolve whether the
``$\Lambda_b$ lifetime problem" is a theoretical or an experimental one (i.e.,
theory predicts $\tau_{\Lambda_b}/\tau_{B_s} \sim 0.9$, while the world average
is about 0.8, except a recent CDF measurement giving a ratio near 1).

To discuss possible NP contributions, we concentrate on NP in $\Delta F=2$
processes and assume that (i)~the $3\times3$ CKM matrix is unitary and (ii)
tree-level decays are SM dominated~\cite{Soares:1992xi}.  Then there are two new
parameters for each meson mixing amplitude
\beq\label{NPparam}
M_{12} =\, \underbrace{M_{12}^{\rm SM}\, r_s^2\, e^{2i\theta_s}}_
  {\hspace*{-.5cm}{\rm easy~to~relate~to~data}\hspace*{-.5cm}} 
\,\equiv\, \underbrace{M_{12}^{\rm SM} 
  \big(1 + h_s\, e^{2i\sigma_s}\big)}_{\rm easy~to~relate~to~NP} .
\eeq
We use the $h, \sigma$ parameterization, since any NP model would give an
additive contribution to $M_{12}$.  To constrain $h$ and $\sigma$, the
measurements of $|V_{ub}/V_{cb}|$ and $\gamma$ (or $\pi-\beta-\alpha$) that
come from tree-level processes and are therefore unaffected by the NP are
crucial~\cite{Ligeti:2004ak}.  One can then compare these with the $B\Bbar$
mixing dependent observables sensitive to  $h$ and $\sigma$, which include
$\Delta m_{d,s}$, $S_{f_i}$, $A_{\rm SL}^{d,s}$, $\Delta\Gamma_s^{CP}$. (As
mentioned above, the hadronic uncertainties are sizable in $A_{\rm SL}^{d,s}$
and  $\Delta\Gamma_s^{CP}$, but in the SM $A_{\rm SL}^{d,s} \ll$ current bound,
while for $\Delta\Gamma_s^{CP}$ they are comparable.  If hadronic uncertainties
are treated conservatively, improving the measurement of $\Delta\Gamma_s^{CP}$
will not yield a better constraint unless LQCD determines the bag parameters
with smaller errors, while the bound from $A_{\rm SL}^{d,s}$ will improve 
independent of progress in LQCD.)

\begin{figure*}[tb]
\includegraphics*[width=.9\columnwidth]{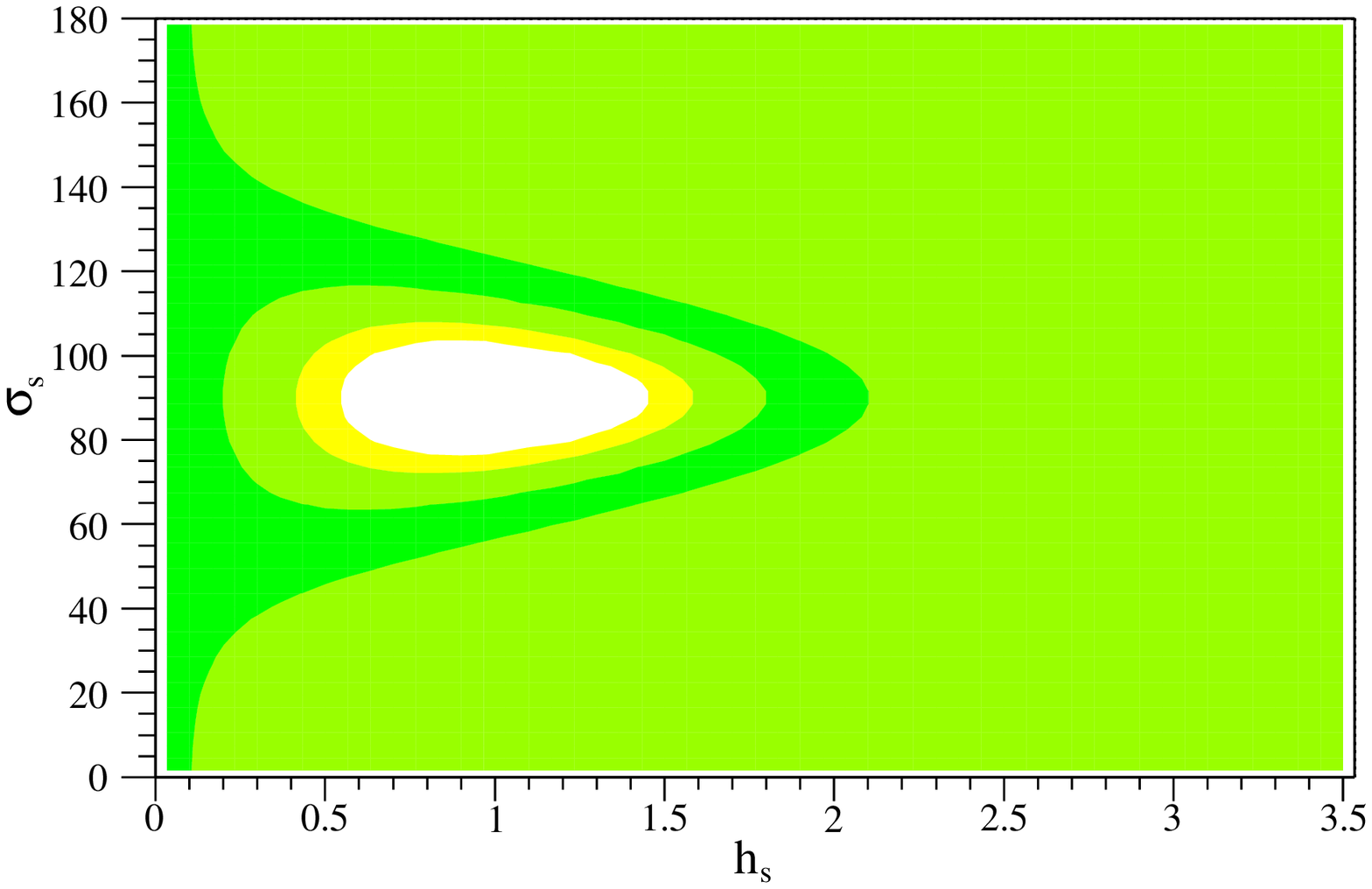} \hfil
\includegraphics*[width=.9\columnwidth]{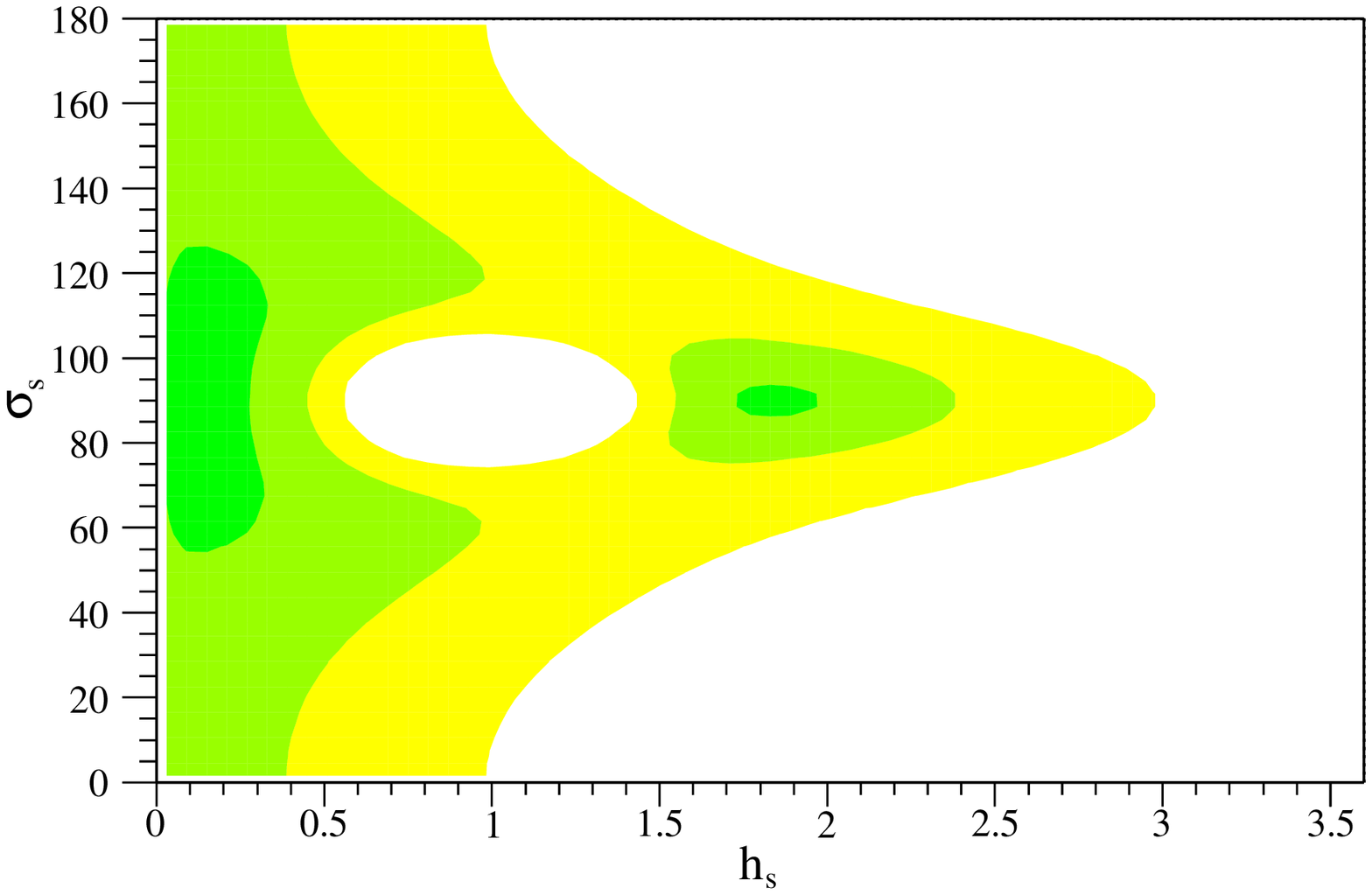}
\\[6pt]
\includegraphics*[width=.9\columnwidth]{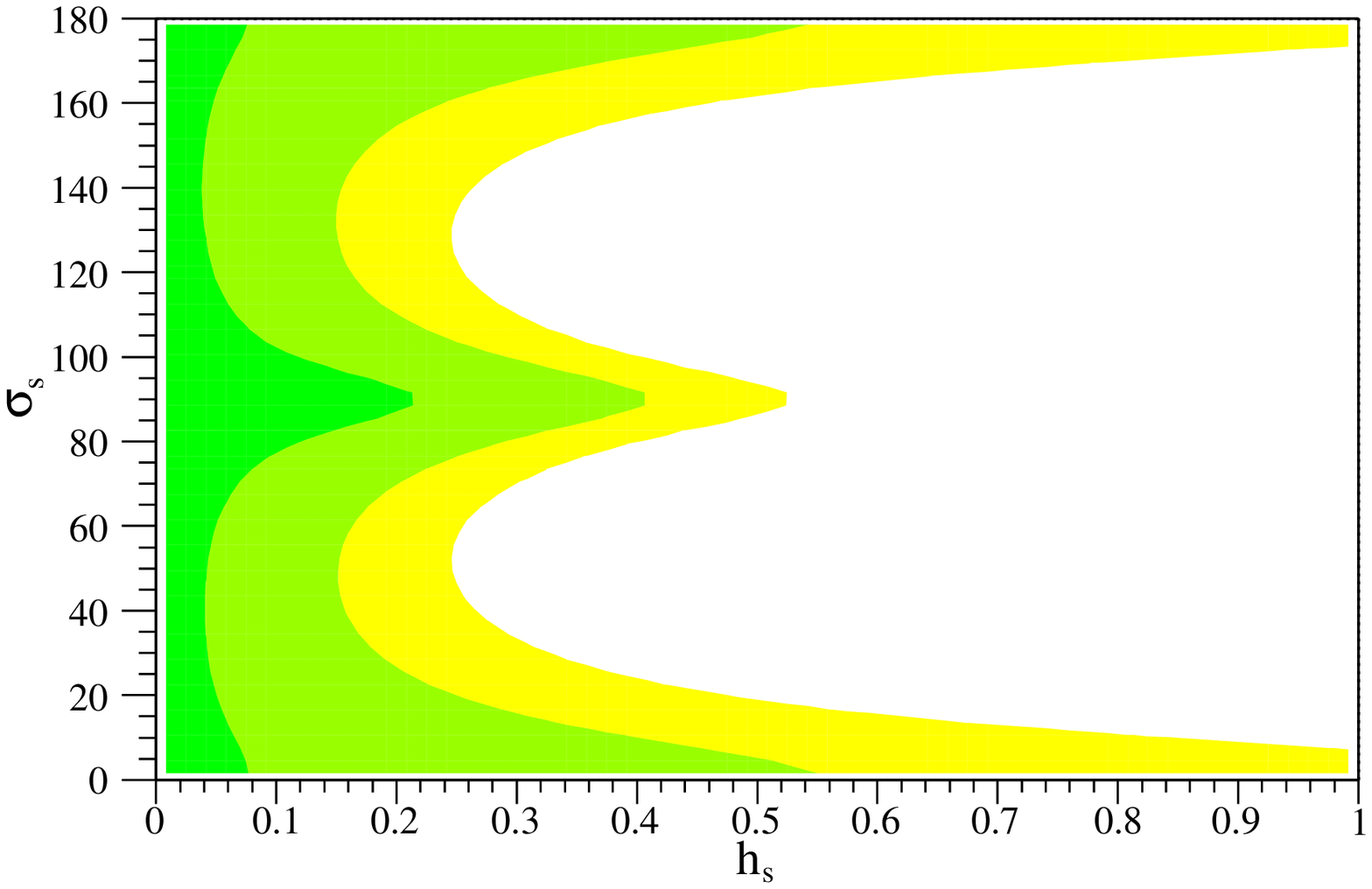} \hfil
\includegraphics*[width=.9\columnwidth]{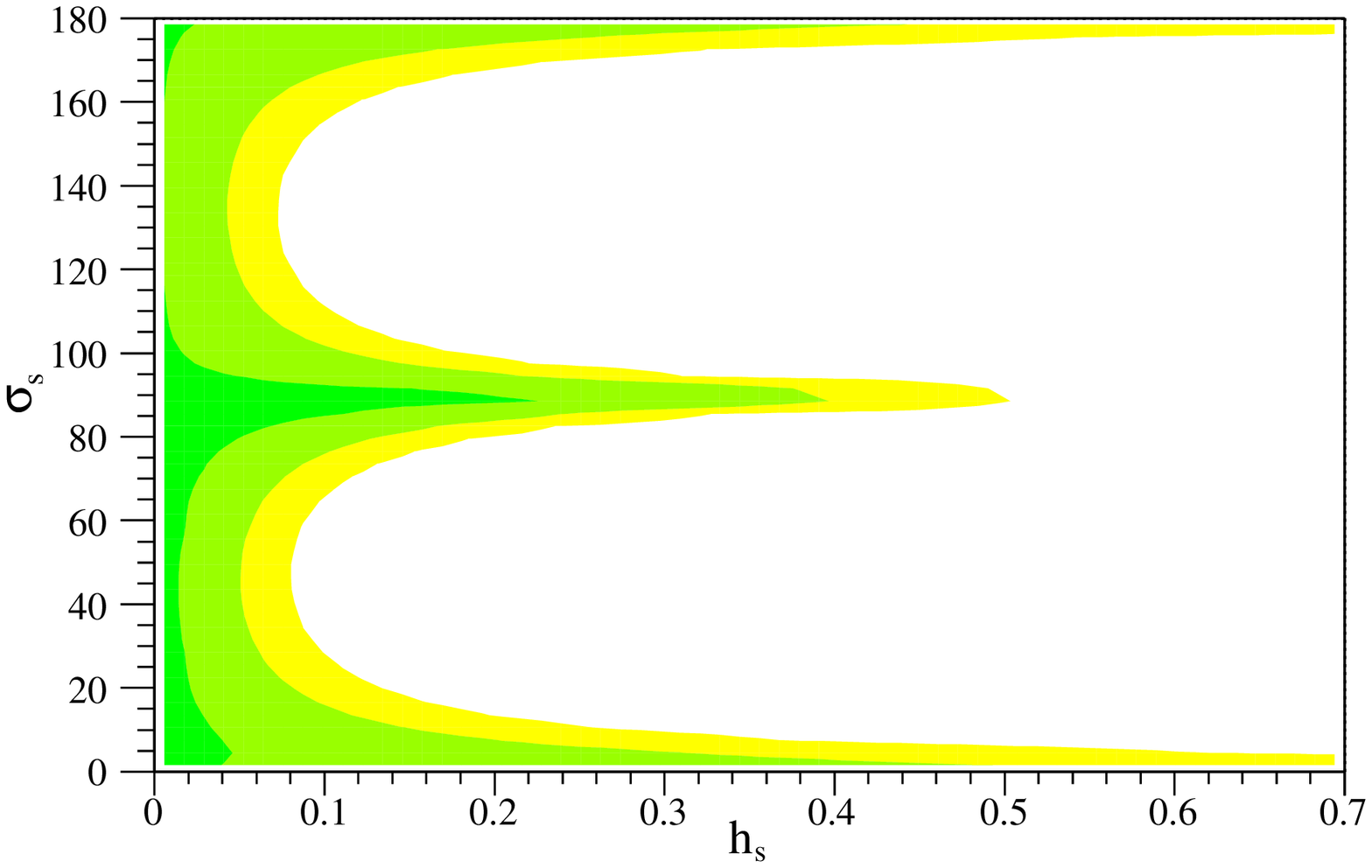}
\caption{Constraints on $h_s$ and $\sigma_s$ before (top left) and after (top
right) the $\Delta m_s$ and $\Delta\Gamma_s^{CP}$ measurements.  The second row
(note the different scales) shows the impact of a future masurement of
$S_{\psi\phi}$ with $\sigma(S_{\psi\phi}) = 0.1$ (bottom left) and $0.03$
(bottom right), expected after 0.1\,yr and 1\,yr of nominal LHCb data.  The dark
green, light green, and yellow areas have CL $>$ 0.90, 0.32, and 0.05,
respectively, indicating the theory uncertainty, $1\sigma$, and $2\sigma$
regions (from~\cite{Ligeti:2006pm}).}
\label{fig:lhcbs}
\end{figure*}

The NP parameters modify the SM predictions as
\beqa\label{par}
\Delta m_s &=& \Delta m_s^{\rm SM}\,
  \big|1+h_q e^{2i\sigma_q}\big| , \nn\\
\Delta\Gamma_s^{CP} &=& |\Delta\Gamma_s^{\rm SM}|\,
  \cos^2\! \big[\arg\big(1+h_s e^{2i\sigma_s}\big)\big] .
\eeqa
The top row in Fig.~\ref{fig:lhcbs} shows the constraint on $h_s$ and $\sigma_s$
before (left) and after (right) the Tevatron measurements of $\Delta m_s$ and
$\Delta\Gamma_s^{CP}$.  To further restrict the parameter space, one needs
measurements sensitive to the $CP$ violating phase in $B_s$ mixing, which will
come from $S_{\psi\phi}$, the time dependent $CP$ asymmetry in $B_s\to
J/\psi\,\phi$.  This is the analog of $S_{\psi K} = \sin2\beta$ in $B_d \to
J/\psi K_S$.  In the SM, $S_{\psi\phi} = \sin2\beta_s$ for the $CP$-even part of
the final state, where
\beq
\beta_s  = \arg( - V_{ts}V_{tb}^* / V_{cs}V_{cb}^*) = {\cal O}(\lambda^2)\,,
\eeq
is the small angle in one of the ``squashed" unitarity triangles, for which the
CKM fit predicts $\sin2\beta_s = 0.0365^{+0.0021}_{-0.0020}$~\cite{ckmfitter}. 
In the presence of NP
\beq
S_{\psi\phi} = \sin\! \big[2\beta_s - \arg \big(1+h_s e^{2i\sigma_s}\big)\big] .
\eeq
Just like when the first $B$ factory results emerged in 2000 the first question
was whether $\sin2\beta$ was consistent with the constraints at that time
(mainly from $\epsilon_K$, $|V_{ub}/V_{cb}|$, and $\Delta m_B$), in 2009 the
question will be if the first measurements of $\sin2\beta_s$ are consistent with
its smallness predicted by the SM.  It is not necessary to measure it with a
sensitivity near the SM to make a significant impact, and CDF or D\O\ may also
be able to do a first measurement~\cite{Anikeev:2001rk,Abazov:2007zj}. 
Observing a sizable nonzero value of $S_{\psi\phi}$ would disprove both the SM
and minimal flavor violation (MFV) scenarios.

The plots in the second row in Fig.~\ref{fig:lhcbs} show the constraints on
$h_s$ and $\sigma_s$ when the measurement of $S_{\psi\phi}$ will be available
with an error of 0.1 (left) and 0.03 (right), which are expected with 0.1 and 1
year of nominal LHCb data.  Such a relatively small data set will constrain
$h_s$ below $0.1$, except if $\sigma_s$ is near 0 (mod $\pi/2$), where
significant deviations from the SM will still be allowed, but only in a way
consistent with MFV.  These two plots do not contain a constraint from
$\Delta\Gamma_s^{CP}$, which may be dominated by hadronic uncertainties by that
time.

The parameter $h$ gives some measure of ``fine tuning".  We expect generically
$h\sim (4\pi v/\Lambda)^2$, so as long as $h \sim 1$ is allowed, the flavor
scale can be $\Lambda_{\rm flavor} \sim 2\,\TeV \sim \Lambda_{\rm EWSB}$, while
if future data constrain $h < 0.1$ then $\Lambda_{\rm flavor} > 7\,\TeV \gg
\Lambda_{\rm EWSB}$.  If NP is seen at the LHC and the constraints on the flavor
scale are pushed up near 10\,\TeV, i.e., if $h < 0.1$ can be achieved, we shall
know that some additional mechanism is present suppressing FCNC's.

\begin{figure}[bth]
\includegraphics*[width=.9\columnwidth]{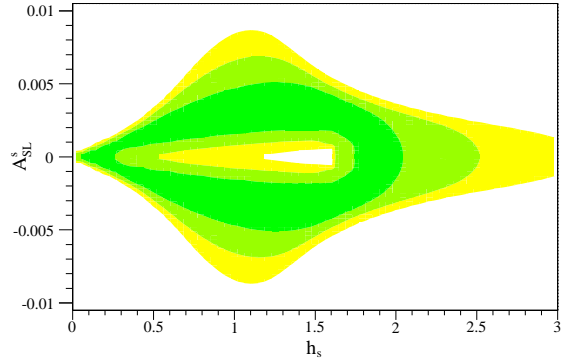}
\caption{Allowed region of $A_{\rm SL}^s$ and $h_s$
(from~\cite{Ligeti:2006pm}).}
\label{fig:hsasls}
\end{figure}

Another interesting observable which can constrain NP~\cite{Laplace:2002ik}, and
has recently been started to be constrained experimentally is $A_{\rm
SL}^{d,s}$,
\beqa
A_{\rm SL}(t)
&=& {\Gamma[\Bbar^0(t)\to\ell^+X] - \Gamma[B^0(t)\to\ell^-X]\over
  \Gamma[\Bbar^0(t)\to\ell^+X] + \Gamma[B^0(t)\to\ell^-X]} \nn\\
&=& {1-|q/p|^4\over 1+|q/p|^4}
  \approx 2 \big(1 - |q / p| \big) ,
\eeqa
which is actually time-independent, and measures the difference between the
$B\to \Bbar$ and $\Bbar\to B$ probabilities~\cite{old}.  In the SM, $A_{\rm
SL}^s \sim 3 \times 10^{-5}$~\cite{Beneke:2003az} is unobservably small. In $K$
decay the similar asymmetry has been measured~\cite{cplear}, in agreement with
the expectation that it is $4\,{\rm Re}\, \epsilon$.  In the presence of 
NP~\cite{Ligeti:2006pm,Blanke:2006ig,Grossman:2006ce}
\beq
A^s_{\rm SL} = {\rm Im} \big\{ \Gamma_{12}^s / \big[ M_{12}^{s,{\rm SM}}
  (1+h_s e^{2i\sigma_s}) \big] \big\} .
\eeq
Figure~\ref{fig:hsasls} shows the allowed region of $A^s_{\rm SL}$ as a function
of $h_s$.  Interestingly, $A^s_{\rm SL}$ can still be as much as  ${\cal
O}(10^3)$ times its SM value, and $|A_{\rm SL}^s| > |A_{\rm SL}^d|$ is possible,
contrary to the SM.

\begin{figure}[tb]
\includegraphics*[width=.9\columnwidth]{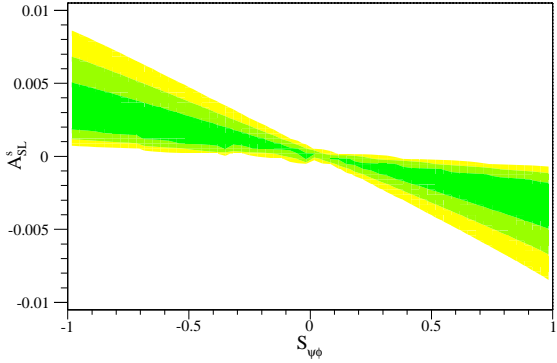}
\caption{Correlation between $A_{\rm SL}^s$ and $S_{\psi\phi}$ 
(from~\cite{Ligeti:2006pm}).}
\label{fig:aslcorr}
\end{figure}

Due to the smallness of $\beta_s$ in the SM, $A_{\rm SL}^s$ and $S_{\psi\phi}$
are strongly correlated in the region of NP parameter space in which
$h_s,\sigma_s\gg \beta_s$~\cite{Ligeti:2006pm}
\beq
A_{\rm SL}^s = - \left|\Gamma^s_{12}\over M^s_{12}\right|^{\rm SM} 
  S_{\psi\phi} + {\cal O}\bigg(h_s^2,\, {m_c^2\over m_b^2}\bigg) .
\eeq
This correlation, which holds in any model where NP does not affect tree level
processes, is plotted in Fig.~\ref{fig:aslcorr}, including theoretical
uncertainties.  Should the measured values violate this correlation, we would
know that NP cannot be parameterized simply by Eq.~(\ref{NPparam}).

\subsection{\boldmath Summary for $B_s^0$\,--\,$\B0bar_s$ mixing}

\begin{itemize}\itemsep 0pt

\item Measurements at the Tevatron started to constrain new physics in $\Delta
F=2$ $b\to s$ transitions.

\item Nevertheless, significant non-SM contributions are still allowed.

\item To make progress, measurements of $S_{\psi\phi}$ and $A_{\rm SL}^s$ are
needed (but sensitivity at the SM level is not required to have important
implications).

\item LHCb can distinguish between MFV and non-MFV scenarios (observation of
$S_{\psi\phi} \neq 0$ at the Tevatron would rule out the SM and MFV).

\item If evidence for NP is found, the correlation of $S_{\psi\phi}$ and $A_{\rm
SL}^s$ may help to understand its nature.

\end{itemize}\vspace*{-6pt}

\begin{table*}[t!]
\caption{SM expectations and measurements of $x$ and $y$ (neglecting $CP$
violation in  mixing), and $(1 - |q/p|^4)/(1 + |q/p|^4)$.}
\label{tab:xyA}
\centerline{\tabcolsep 4pt
\begin{tabular}{c|cc|cc|cc} \hline\hline
  &  \multicolumn{2}{c|}{$x = \Delta m/\Gamma$}
  &  \multicolumn{2}{c|}{$y = \Delta\Gamma/(2\Gamma)$}
  &  \multicolumn{2}{c}{$A = (1 - |q/p|^4)/(1 + |q/p|^4)$}  \\[-4pt]
  &  SM theory  &  data  &  SM theory  &  data  &  SM theory  &  data \\
\hline\hline
$B_d$  &  ${\cal O}(1)$  &  $0.775$  &  $y_s\, |V_{td}/V_{ts}|^2$  &
  $-0.005\pm0.019$  &  $-(5.5\pm1.5) 10^{-4}$  &  $(-4.7\pm4.6)\!\times\!10^{-3}$ \\
$B_s$  &  $x_d\, |V_{ts}/V_{td}|^2$  &  $25.8$  &  ${\cal O}(-0.1)$  &
  $-0.05\pm0.04$  &  $-A_d\,|V_{td}/V_{ts}|^2$  &  $(0.3\pm9.3)\!\times\!10^{-3}$\\ \hline
$K$  &  ${\cal O}(1)$  &  $0.948$  &  $-1$  &  $-0.998$  &
  $4\, {\rm Re}\, \epsilon$  &  $(6.6\pm1.6)\!\times\!10^{-3}$ \\ \hline
$D$  &  $\lsim 0.01$  &  $<0.016$ (95\%\,CL)  &  ${\cal O}(0.01)$  &
  $0.011\pm0.003$ ($y_{CP}$)  &  $<10^{-4}$  &  ${\cal O}(0.5)$ bound only \\ 
\hline\hline
\end{tabular}}
\end{table*}

\section{Concluding remarks}

Instead of a usual summary, Table~\ref{tab:xyA} shows the SM predictions and the
current experimental information on the mixing parameters, $x = \Delta
m/\Gamma$, $y = \Delta\Gamma/(2\Gamma)$, and $A = (1 - |q/p|^4)/(1 + |q/p|^4)$. 
While $|q/p|$ is very near 1 in the $K^0$, $B_d^0$, and $B_s^0$ systems, we do
not know this for the $D^0$ yet (it does hold in the SM).  The correspondence
between the lifetimes, $CP$ eigenstates, and mass eigenstates of the neutral
mesons, in the limit neglecting $CP$ violation, is
\beqa\label{identity}
K \!\!&:& \mbox{long-lived} = CP\mbox{-odd} = \mbox{heavy} , \\
D \!&:& \mbox{long-lived} = CP\mbox{-odd } (3.5\sigma)
  = \mbox{light } (2\sigma) , \nn\\
B_s \!\!&:& \mbox{long-lived} = CP\mbox{-odd } (1.5\sigma)
  = \mbox{heavy in the SM} , \nn\\
B_d \!\!&:& \mbox{yet unknown; same as $B_s$ in SM for 
  $m_b \!\gg\! \Lambda_{\rm QCD}$}. \nn
\eeqa
Taking Belle's $D\to K_S\pi^+\pi^-$ analysis as evidence for the sign of $x/y$
implies that the $CP$-odd $D^0$ state is the lighter one, contrary to the $K^0$
system (and probably the $B_{d,s}$ systems as well).  This information is more
amusing than useful, since it does not tell us which measurements give clean
short-distance information.~\,Curiously, before 2006 we only knew experimentally
the first line in (\ref{identity}).

As an aside, note that in the $B_d^0$ system it is hard, if not impossible, to
identify the $CP$-even and odd states simply by their decays to $CP$
eigenstates.  Although $B_{L,H}$ can be defined as almost pure $CP$ eigenstates,
both $B_{L,H}$ can decay to the same $CP$ eigenstates, since the weak
interaction responsible for the decays does not conserve $CP$.\footnote{I thank
Klaus Schubert for emphasizing this point to me.}  If the phase of the decay and
the mixing amplitudes are not the same ($V_{tb}V_{td}^*$), i.e., if $\lambda
\neq \pm 1$, then the untagged $B$ decay rate is
\beqa\label{CPdecay}
\Gamma(B\to f) &\propto& 
  \left(1+{2\,{\rm Re}\,\lambda\over 1+|\lambda|^2}\right) e^{-\Gamma_L t} \nn\\
&+& \left(1-{2\,{\rm Re}\,\lambda\over1+|\lambda|^2}\right) e^{-\Gamma_H t} \,,
\eeqa
indicating that both $B_H$ and $B_L$ can decay to the same final $CP$
eigenstate.  It is not yet known if $CP$ violation is absent in any decay to a
$CP$ eigenstate.  It would be if $b\to d$ penguins (e.g., $B\to \phi\pi^0$) were
dominated by the top loop, however, the $V_{tb}V_{td}^*$ and $V_{cb}V_{cd}^*$
terms are comparable.  The best hope, in principle, may be $B\to \rho^+\rho^-$,
if the data converge toward $\alpha$ near $90^\circ$ and small penguin to tree
ratio.

Looking into the future, some of the most interesting measurements which I hope
will emerge are as follows.
In $D^0$\,--\,$\D0bar$ mixing:
\begin{itemize}\vspace*{-4pt}\itemsep 0pt
\item More robust measurements of $\Delta m$ and $\Delta\Gamma$;
\item Will CPV be observed?\, Is $|q/p|$ near 1?
\item Result of $K\pi$ fit with $5$ parameters (allowing $CP$ violation 
in mixing, but not in decay).
\end{itemize}\vspace*{-4pt}
In $B_s^0$\,--\,$\B0bar_s$ mixing:
\begin{itemize}\vspace*{-4pt}\itemsep 0pt
\item Better constraint on / measurement of $S_{\psi\phi}$;
\item Improved bounds on $A_{\rm SL}$;
\item Better lattice QCD results for $\Delta m$ and $\Delta\Gamma$.
\end{itemize}\vspace*{-4pt}
Clearly, we can learn a lot from these measurements, so it will be exciting to
see what they teach us over the next several years.  Either new physics signals
may be observed, or the flavor structure of the SM will have been tested (or
that of the NP seen at the LHC constrained) at a whole new level, providing
insights to the physics of flavor changing interactions.

\begin{acknowledgments}

I thank Bob Cahn, Bostjan Golob, Yuval Grossman, Yossi Nir, and Klaus Schubert
for enjoyable discussions.
I thank Peter Krizan and Bostjan Golob for the invitation and for organizing a
delightful conference.
This work was supported in part by the Director, Office of Science, Office
of High Energy Physics of the U.S.\ Department of Energy under the
Contract DE-AC02-05CH11231.
\end{acknowledgments}

\end{document}